\documentclass[12pt]{article}
\usepackage{amsfonts}

\usepackage{pstricks}
\usepackage{pst-node}
\usepackage{epsfig}
\usepackage{amsmath,amssymb,latexsym}
\usepackage{xcolor}
\usepackage{hyperref}
\usepackage{slashed} 

\usepackage{import}

\usepackage{enumerate}

\usepackage{physics}

\usepackage[lmargin=1.0in, rmargin=1.0in, tmargin=1.0in, bmargin=1.0in]{geometry}

\usepackage{cancel}

\usepackage{empheq}

\usepackage{comment}

\usepackage{setspace}
\usepackage{ulem}

\begin{document}



\title{\bf Generalized Dirac Equation in Metric-Affine Spacetimes:
Consistency, Effective Mass, and Spinor Curvature}
\author{Muzaffer Adak$^1$, Ali Bagci$^1$, Caglar Pala$^2$, Ozcan Sert$^1$  \\
  {\small $^1$Computational and Gravitational Physics Laboratory, Department of Physics,} \\
   {\small Faculty of Sciences, Pamukkale University, Denizli, Türkiye} \\
  {\small $^2$Department of Physics,
Faculty of Arts and Sciences, Erciyes University, Kayseri, Türkiye}\\
      {\small {\it E-mail:} {\blue madak@pau.edu.tr, abagci@pau.edu.tr, caglar.pala@gmail.com, osert@pau.edu.tr}}}

  \vskip 1cm
\date{\today}
\maketitle
\thispagestyle{empty}
\begin{abstract}
 \noindent
We formulate a generalized Dirac equation in four-dimensional
metric-affine spacetimes with curvature, torsion, and non-metricity.
The spinor connection is expanded over the complete basis of the
Clifford algebra $cl(1,3)$ and contains eight complex coupling
constants. Requiring agreement between direct covariantization
of the Dirac equation and variation of the corresponding
symmetrized Lagrangian imposes eight real consistency conditions,
reducing the number of independent real parameters from sixteen
to eight. In the static isotropic symmetric teleparallel sector,
formal Hermiticity of the Hamiltonian with respect to the
curved-space Dirac inner product reproduces the corresponding
restrictions on the non-metricity and coframe couplings.
The coframe-dependent terms introduce scalar and pseudoscalar
mass contributions that persist in Minkowski spacetime.
For a single free Dirac field with constant coefficients,
they yield one effective mass,
$m_{\mathrm{eff}}=\sqrt{(m-4b_3)^2+16b_4^2}$,
and a chiral phase removable by a constant chiral field redefinition.
A nonzero effective mass is therefore possible even when
the original mass parameter vanishes. We also derive the
Clifford decomposition of the generalized spinor curvature
and show that it can remain nonzero when affine curvature,
torsion, and non-metricity all vanish. When $m=0$, the
combination $b_3^2+b_4^2$ determines both the effective mass
squared and the surviving spinor curvature, establishing
an algebraic link between these contributions.
\\


 \noindent
 {\it Keywords}: Curvature, torsion, non-metricity, metric-affine spacetimes, Dirac equation, spinor bundle, spinor connection.

\end{abstract}

\section{Introduction} \label{sec:intro}

The Dirac equation provides the relativistic description of spin-$1/2$
particles and a starting point for the treatment of fermionic fields
\cite{dirac1928,bjorken1964,greiner2000,simulik2025}. Coupling a Dirac
field to gravity requires an orthonormal coframe and a spinor connection
that specifies its covariant derivative
\cite{fock1929,de_oliveira1962,arminjon2008,alcubierre2025}.
In the language of exterior algebra, the usual minimal coupling
prescription is expressed by the replacement $d\rightarrow D$, with
the coframe and the spin connection encoding the gravitational
background \cite{dereli-tucker1982}. In a torsion-free,
metric-compatible spacetime, the Levi-Civita spin connection yields
the standard curved-spacetime Dirac equation, and applying this
prescription to the field equation or to the symmetrized Dirac
Lagrangian gives equivalent results.

Metric-affine geometry extends this setting by treating the metric
and the affine connection as independent structures and allowing
both torsion and non-metricity. Einstein--Cartan theory provides
a familiar example in which the connection is metric compatible
but torsion is sourced algebraically by the spin density of matter
\cite{trautman1973,trautman2006}. Eliminating torsion in the coupled
Einstein--Cartan--Dirac system produces effective fermionic
self-interactions and cubic terms in the Dirac equation
\cite{gursey1957,khanapurkar2018}. The coupling of spinor matter
to such geometries also requires care in relating the
connection, the variational principle, and the resulting field
equations \cite{hehl-datta1971,obukhov-pereira2004}. In particular,
the direct replacement of the ordinary derivative in the free
Dirac equation need not reproduce the equation obtained by varying
the corresponding covariantized Lagrangian
\cite{delhom2020,adak2012,adak-ozdemir2023}. Integration by parts
in a non-Riemannian geometry introduces additional geometric
terms, making agreement between the direct and variational
formulations a nontrivial requirement.

Non-metricity introduces further questions concerning parallel
transport and the relation between spacetime geometry and matter
couplings \cite{ponomarev1982,adak2023sce,vacaru2025}.
Although the Clifford algebra is well defined locally by the
metric in an orthonormal frame, specifying how the general affine
connection acts on Dirac spinors requires a prescription beyond
the conventional Lorentz spin connection. Approaches to this
problem include extended spinor representations, world spinors,
nonlinear connection structures, and generalized spinor
covariant derivatives
\cite{neeman1977,vacaru2004,adak2003,jimenez2020,hehl1995}.
These considerations motivate a formulation in which the
geometric ingredients of the spinor connection and the
conditions required for variational consistency are stated
explicitly.

In this work, we construct a generalized spinor covariant
derivative $\mathbb{D}\psi=(d+\Omega)\psi$ by expanding the
connection over the complete basis of the Clifford algebra
$cl(1,3)$. The proposed connection combines the antisymmetric
part of the affine connection with the two traces of
non-metricity, the torsion trace, and coframe-dependent terms.
Its coefficients comprise eight arbitrary complex constants,
equivalently sixteen real parameters. We derive the generalized
Dirac equation both directly and variationally and require the
two equations to coincide. This requirement imposes eight
independent real constraints, leaving eight real coupling
parameters. It therefore supplies an internal consistency
criterion for selecting admissible connections within the
proposed framework.

We also examine the physical meaning of these constraints
through the Hermiticity of the generalized Dirac Hamiltonian.
In the static isotropic symmetric teleparallel sector studied
in Ref.~\cite{ugur_adak_bagci_pala_2026}, formal Hermiticity,
evaluated using the curved-space Dirac inner product and the
complete position-dependent kinetic operator, reproduces the
corresponding restrictions on the non-metricity and coframe
couplings. This agreement connects direct--variational
consistency with probability-conserving fermionic evolution.
Since torsion vanishes in this sector, the Hamiltonian analysis
does not independently constrain the torsion couplings.

The coframe-dependent coefficients $b_3$ and $b_4$ also extend
the fermionic mass sector. They introduce scalar and
pseudoscalar contributions through the mass operator
$\mathsf{M}=(m-4b_3)I-4b_4\gamma_5$, which remains present
when the underlying affine geometry is Minkowski spacetime.
For a single free Dirac field with constant coefficients,
the dispersion relation contains one effective mass,
$m_{\mathrm{eff}}=\sqrt{(m-4b_3)^2+16b_4^2}$.
The chiral coefficients are complex conjugates, and their
relative phase can be removed by a constant chiral field
redefinition in this free limit. Thus, the extended mass
structure involves a single rest mass and a chiral phase.
In particular, a nonzero effective mass is possible even
when the original mass parameter vanishes. The physical
significance of the scalar--pseudoscalar decomposition in
an interacting theory must be assessed by transforming
all interaction terms consistently.

Finally, we compute the endomorphism-valued curvature
$\mathcal{R}=d\Omega+\Omega\wedge\Omega$ of the generalized
spinor connection. The Clifford decomposition of this
curvature distinguishes it from the curvature of the
underlying affine connection. The noncommuting
coframe-dependent terms can produce a nonzero spinor
curvature even when affine curvature, torsion, and
non-metricity all vanish. When $m=0$, the same combination
$b_3^2+b_4^2$ controls both the effective mass squared and
the surviving spinor curvature. This correspondence
relates the mass and curvature contributions carried
by the generalized connection while the underlying
affine spacetime remains flat.

The paper is organized as follows.
Section~\ref{sec:math-preli} introduces metric-affine
geometry and the exterior-algebra identities used in
the calculations.
Section~\ref{sec:spinor-bundle} reviews the Clifford
algebra and the spinor bundle.
Section~\ref{sec:evolution-of-dirac-eqn} compares the
direct and variational Dirac equations in Minkowski,
Riemannian, and Riemann--Cartan spacetimes.
Section~\ref{sec:our-solution-to-inconsis} develops the
generalized spinor connection, derives the consistency
conditions, and discusses their relation to Hamiltonian
Hermiticity.
Section~\ref{sec:effective mass} analyzes the effective
mass and chiral structure, including the Minkowski limit,
chiral field redefinitions, and the equivalence of the
symmetrized and unsymmetrized Lagrangians up to an exact
four-form.
Section~\ref{sec:spinor-curvature} derives the curvature
of the generalized spinor connection and examines its
limiting cases.
Section~\ref{sec:discussion} presents the conclusions.

\section{Metric-affine geometry} \label{sec:math-preli}

The triple $\{M,g,\nabla\}$ or $\{Q_{ab},T^a,R^a{}_b\}$ defines a metric-affine geometry where $M$ is a four-dimensional orientable and differentiable manifold, $g$ is a non-degenerate symmetric metric, $\nabla$ is a full (or affine) connection or covariant derivative, $Q_{ab}$ is a non-metricity 1-form, $T^a$ is a torsion 2-form, and $R^a{}_b$ is a curvature 2-form \cite{thirring1997,frankel2012}. We denote the anholonomic metric-orthonormal coframe by $e^a$, and write the metric as $g=\eta_{ab} e^a \otimes e^b$ where $\eta_{ab}$ is the Minkowski metric with the signature $(-,+,+,+)$. Latin indices are the anholonomic orthonormal (Lorentz) spacetime indices; $a,b,c, \cdots = 0,1,2,3$. The affine connection is determined by the full connection 1-form $\omega^a{}_b$ by the definitions $\nabla e^a := - \omega^a{}_b \wedge e^b$ or $\nabla X_a := X_b \otimes \omega^b{}_a$ where $\otimes$ and $\wedge$ denote the tensor product and exterior product, respectively, and $X_a$ is the orthonormal frame such that $e^a(X_b)=\delta^a_b$ is the duality relation. Besides, $e^a$ is called the metric orthonormal (orthonormal, for short) basis 1-form, and the Cartan structure equations are given by the tensor-valued non-metricity 1-form, the tensor-valued torsion 2-form, and the tensor-valued full (or non-Riemannian) curvature 2-form, respectively,
 \begin{subequations}\label{eq:cartan-ort}
 \begin{align}
     Q_{ab} &:= -\frac{1}{2} D\eta_{ab}  = -\frac{1}{2} \left( d\eta_{ab} - \omega^c{}_a \eta_{cb} - \omega^c{}_b \eta_{ac} \right) = \omega_{(ab)} , \label{eq:nonmetric}\\
     T^a &:= De^a = de^a + \omega^a{}_b \wedge e^b, \label{eq:tors}\\
     R^a{}_b &:= D\omega^a{}_b := d \omega^a{}_b + \omega^a{}_c \wedge \omega^c{}_b, \label{eq:curv}
 \end{align}
 \end{subequations}
where $d$ is the exterior derivative and $D$ is the covariant exterior derivative\footnote{Here $\omega_{(ab)} = Q_{ab}$ is the reason why we adopt the factor $-\frac{1}{2}$ in the definition of non-metricity}. We used the result $d\eta_{ab}=0$ in the orthonormal frame given by Eq.~(\ref{eq:nonmetric}). We utilize the usual notation: $( \bullet )$ and $[ \bullet ]$ mean to calculate the symmetric and anti-symmetric parts, respectively, with respect to the indices enclosed, i.e., $  \omega_{(ab)} = \frac{1}{2} (\omega_{ab} + \omega_{ba})$ and $  \omega_{[ab]} = \frac{1}{2} (\omega_{ab} - \omega_{ba})$. They satisfy the Bianchi identities
  \begin{align} \label{eq:bianchi_identities}
      DQ_{ab} = R_{(ab)} , \qquad 
      DT^a = R^a{}_b \wedge e^b , \qquad
      DR^a{}_b = 0 .
  \end{align}

If needed, one can easily switch to tensor notation by expanding the corresponding exterior forms in components as follows
  \begin{align}
     \omega^a{}_b = \omega^a{}_{bc} e^c , \qquad  Q_{ab}=Q_{abc} e^c , \qquad T^a = \frac{1}{2} T^a{}_{bc} e^b \wedge e^c , \qquad R^a{}_b = \frac{1}{2} R^a{}_{bcd} e^c \wedge e^d .
  \end{align}
Thus, we deduce that $Q_{abc}=Q_{(ab)c}$ is symmetric in its first two indices, $T^a{}_{bc}=T^a{}_{[bc]}$ is antisymmetric in its last two indices, and $R_{abcd} = R_{ab[cd]}$ is antisymmetric in its last two indices and asymmetric $R_{abcd} = R_{[ab]cd} + R_{(ab)cd}$ at the first two indices by definition. In the subsequent sections, the definitions of the first kind trace 1-form $Q :=\eta_{ab} Q^{ab} = Q^a{}_{ab} e^b$ and the second kind trace 1-form $P :=(\iota_aQ^{ab}) e_b = Q^a{}_{ba} e^b$ of the non-metricity tensor, as well as the trace 1-form of the torsion $T:=\iota_a T^a$ will be frequently used, where $\iota_a \equiv \iota_{X_a}$ denotes the interior derivative, $\iota_b e^a = \delta^a_b$. Furthermore, we would like to explicitly remind you that $d\eta_{ab}=0$, $d\eta^{ab}=0$, and $d\delta^a_b=0$, but $D\eta_{ab} = -2Q_{ab}$, $D\eta^{ab}=+2Q^{ab}$, and $D\delta^a_b=0$. Accordingly, while one may freely raise or lower an orthonormal index in front of the $d$ operator, one must pay special attention to raising or lowering an index in front of the $D$ operator in non-metric geometries. Furthermore, the following identities are very useful for the calculations. 
 \begin{subequations}
     \begin{align}
          D *e_a &=  -Q \wedge *e_a + T^b \wedge *e_{ab} \label{eq:Dhodgeea} \\
          D *e_{ab} &=  -Q \wedge *e_{ab} + T^c \wedge *e_{abc} \\
          D *e_{abc} &=  -Q \wedge *e_{abc} + T^d \wedge *e_{abcd} \\
          D *e_{abcd} &=  -Q \wedge *e_{abcd} 
     \end{align}
 \end{subequations}


Nevertheless, the full connection 1-form, $\omega^a{}_b$, can be uniquely decomposed into a Riemannian piece, $\widetilde{\omega}_{ab}(g)$, determined by the metric, plus a non-Riemannian piece, $\mathrm{L}_{ab}(T,Q)$ determined by torsion and non-metricity \cite{hehl1995,adak2023ijgmmp,tucker1995}
 \begin{align}
     \omega_{ab}=\widetilde{\omega}_{ab} + \mathrm{L}_{ab}, \label{eq:connec-decom}
 \end{align}
where $\widetilde{\omega}_{ab}=-\widetilde{\omega}_{ba}$ is the Levi-Civita connection 1-form, 
 \begin{equation}\label{eq:Levi-Civita}
   \widetilde{\omega}_{ab} = \frac{1}{2} \left[ -\iota_a de_b + \iota_b de_a + (\iota_a \iota_b de_c) e^c \right] \qquad \text{or} \qquad \widetilde{\omega}^a{}_b \wedge e^b = -de^a
 \end{equation}
and $\mathrm{L}_{ab}$ is the tensor-valued distortion (or defect) 1-form, 
 \begin{equation}
     \mathrm{L}_{ab} := \underbrace{ \underbrace{ \frac{1}{2} \left[ \iota_a T_b - \iota_b T_a - (\iota_a \iota_b T_c) e^c \right]}_{contortion} + \underbrace{ ( \imath_b Q_{ac} - \imath_a Q_{bc} ) e^c + Q_{ab} }_{disformation} }_{distortion}.
 \end{equation}
In the literature, it is common to define the tensor-valued contortion  1-form, $K_{ab}=-K_{ba}$, in terms of the torsion 2-form
 \begin{align} \label{eq:torsion-contor}
     K_{ab} = \frac{1}{2} \left[ \iota_a T_b - \iota_b T_a - (\iota_a \iota_b T_c) e^c \right] \qquad \text{or} \qquad K^a{}_b \wedge e^b = T^a .
 \end{align}
It is important to note that the symmetric part of the affine connection is determined solely by non-metricity, $\omega_{(ab)}= Q_{ab}$. The remainder of $\omega_{ab}$ is the anti-symmetric part, $\omega_{[ab]} = \widetilde{\omega}_{ab} + K_{ab} +  ( \imath_b Q_{ac} - \imath_a Q_{bc} ) e^c$.

When non-metricity is set to zero, the full connection is called metric compatible, in which case the symmetric parts of the full connection and full curvature vanish. If torsion is also set to zero along with non-metricity, the full connection is named the Levi-Civita (or Riemannian) connection. An affine geometry is classified based on whether non-metricity, torsion, and/or full curvature vanish or not \cite{adak2023ijgmmp}.

\section{Spinor bundle} \label{sec:spinor-bundle}

The Clifford algebra $cl(1,3)$ is generated by the Minkowski vector space with the signature $(-,+,+,+)$. The set $\{ I, \gamma_0, \gamma_1, \gamma_2, \gamma_3 \}$ of $4 \times 4$ complex matrices could be the generators of $cl(1,3)$ that satisfy the condition
   \begin{align}
       \gamma_a \gamma_b + \gamma_b \gamma_a = 2 \eta_{ab} I . \label{eq:dirac-matrix-cond}
   \end{align}
Besides, we represent the basis of $cl(1,3)$ by $\{I, \gamma_a , \sigma_{ab}, \gamma_a\gamma_5 , \gamma_5 \}$, where $\sigma_{ab} :=(\gamma_a \gamma_b - \gamma_b \gamma_a)/4$ and $\gamma_{5}  := \gamma_0 \gamma_1 \gamma_2 \gamma_3$. We adhere to the following representations of Dirac matrices
  \begin{align} 
 \gamma_{0}   = \begin{pmatrix}
              -iI_2 & 0\\ 
                0 & iI_2
             \end{pmatrix} , \qquad
 \gamma_{k}  = \begin{pmatrix}
            0 & i\sigma^k \\ 
            -i\sigma^k & 0
             \end{pmatrix} ,
\end{align}
where $i=\sqrt{-1}$ is the imaginary unit, $I_2$ is the two-dimensional unit matrix, and $\sigma^k$  $(k=1,2,3)$ are the Pauli matrices. There are relations between our matrices and Bjorken$\&$Drell matrices \cite{bjorken1964} as $i\gamma_0 = \beta$ and $\gamma_0 \gamma_k = \alpha_k$. The basis elements of $cl(1,3)$ satisfy the following useful properties
       \begin{align}  \label{eq:gamma0-GammaAdag-gamma0}
 \gamma_0 I^\dagger \gamma_0 = -I , \qquad  \gamma_0 \gamma_a^\dagger \gamma_0 =  \gamma_a , \qquad \gamma_0 \sigma_{ab}^\dagger \gamma_0 =  \sigma_{ab} , \nonumber \\
 \gamma_0 (\gamma_a \gamma_5)^\dagger \gamma_0 =  - \gamma_a \gamma_5 , \qquad \gamma_0 \gamma_5^\dagger \gamma_0 =  - \gamma_5 ,
         \end{align}
where $^\dagger$ denotes the Hermitian conjugate. The commutation relations of the basis elements of $cl(1,3)$ are computed as
     \begin{subequations} \label{eq:commut-rel-cl-basis}
        \begin{align}
           [I,I] &= [I,\gamma_a] = [I,\sigma_{ab}] = [I,\gamma_a \gamma_5] = [I,\gamma_5] =0 ,\\
          [\gamma_a, \gamma_b] & = 4 \sigma_{ab} , \\
          [\gamma_a,\sigma_{bc}] & = \eta_{ab} \gamma_c - \eta_{ac} \gamma_b , \\
          [\gamma_a,\gamma_b \gamma_5] & = 2 \eta_{ab} \gamma_5 ,\\
          [\gamma_a ,\gamma_5] & = 2 \gamma_a \gamma_5 ,\\
            [\sigma_{ab},\sigma_{cd}] & =  -\eta_{ac} \sigma_{bd} - \eta_{bd} \sigma_{ac} + \eta_{bc} \sigma_{ad} + \eta_{ad} \sigma_{bc} , \\
          [\sigma_{ab},\gamma_c \gamma_5] & = -\eta_{ac} \gamma_b \gamma_5 + \eta_{bc} \gamma_a \gamma_5 , \\
          [\sigma_{ab} ,\gamma_5] & = 0 , \\
           [\gamma_a \gamma_5,\gamma_b \gamma_5] & = 4 \sigma_{ab} , \\
          [\gamma_a \gamma_5 ,\gamma_5] & = -2 \gamma_a , \\
           [\gamma_5 ,\gamma_5] & = 0 .
      \end{align}
      \end{subequations}


In bundle constructions over an $n$-dimensional manifold $M$, the underlying object is called the base manifold, while the object defined over it is called the bundle $E$. The base manifold may be regarded as a union of open regions, $M = U \cup V \cup \cdots$, and to each point of $M$, there corresponds a fiber in the bundle, which is generally a $k$-dimensional vector space $V^k$. Accordingly, one writes $E = U \times V^k$. A map that assigns to each point of $M$ a point in the fiber is called a section, whereas the inverse map is called a projection. Moreover, the group formed by the transformations relating two points in $E$ is called the structure group of $E$ and is denoted by $\mathcal{G}$~\cite{thirring1997,frankel2012}.

In accordance with the aim of this work, we first summarize the coordinate coframe bundle. On the $(1+3)$-dimensional manifold $M$, through each point pass as many independent curves as the dimension of the manifold, which define a coordinate system $x^\mu$ ($\mu=0,1,2,3$). The differentials of these coordinates naturally define the objects $dx^\mu$, called the coordinate coframe, and each is represented by a point on the fiber. In this context, $dx^\mu$ is a section. This bundle is specifically called the coordinate coframe bundle, $CT^*(M) = U \times \mathbb{R}^{1,3}$, where $U \subset M$ is a region and $\mathbb{R}^{1,3}$ is a four-dimensional real vector space. On the other hand, at the same point of $M$, one may choose another coordinate system $x^{\mu'}$ and construct a new coordinate coframe $dx^{\mu'}$. The new and old coordinate coframes correspond to two distinct points on the same fiber of $CT^*(M)$. Any two coordinate coframes on the same fiber are related by a linear transformation, $dx^{\mu'} = L^{\mu'}{}_\mu dx^\mu$, where the transformation matrix $L^{\mu'}{}_\mu := \partial x^{\mu'}/\partial x^\mu$ is an element of the general linear group $GL(4,\mathbb{R})$. Thus, $GL(4,\mathbb{R})$ is the structure group of $CT^*(M)$.

All the above discussion can be repeated for the coordinate frame $\partial/\partial x^\mu$, which is dual to the coordinate coframe and satisfies $dx^\mu (\partial/\partial x^\nu)=\delta^\mu_\nu$. In this setting, $\partial/\partial x^\nu$ is the projection of this bundle. Just as we relate two coordinate coframes at the same point of $M$ (that is, two points on the same fiber), we also relate coordinate coframes at two distinct points of $M$ (that is, points on different fibers) by means of the connection of the $CT^*(M)$ bundle. Indeed, the connection determines the rule for the parallel transport of the coframe/frame. For the bundle construction to be well-defined, the connection 1-forms at two different points of $CT^*(M)$ are related through the structure group as
$\omega^{\mu'}{}_{\nu'} = L^{\mu'}{}_\mu \omega^\mu{}_\nu L^{\nu}{}_{\nu'} + L^{\mu'}{}_\mu dL^{\mu}{}_{\nu'}$, where $L^{\mu}{}_{\mu'} := \partial x^\mu / \partial x^{\mu'}$ is the inverse of $L^{\mu'}{}_{\mu}$, satisfying $L^\mu{}_{\mu'} L^{\mu'}{}_\nu = \delta^\mu_\nu$. This procedure is referred to as covariantization.

Accordingly, the metric measuring the infinitesimal distance between neighboring points on $M$ is expressed in $CT^*(M)$ as $g=g_{\mu \nu}(x) dx^\mu \otimes dx^\nu = g_{\mu' \nu'}(x') dx^{\mu'} \otimes dx^{\nu'}$, with the components related by $g_{\mu' \nu'} = L^\mu{}_{\mu'} g_{\mu\nu} L^\nu{}_{\nu'}$. Consequently, a $(1,1)$-type tensor-valued exterior form $\mathcal{T}^\mu{}_\nu$ can be defined on $CT^*(M)$, and its covariant exterior derivative is
 \begin{align}
     D \mathcal{T}^\mu{}_\nu = d\mathcal{T}^\mu{}_\nu + \omega^\mu{}_\sigma \wedge \mathcal{T}^\sigma{}_\nu - \omega^\sigma{}_\nu \wedge \mathcal{T}^\mu{}_\sigma \, .
 \end{align}
Under the transformation rule $dx^{\mu'} = L^{\mu'}{}_\mu dx^\mu$, both the tensor components and their covariant exterior derivative transform as follows:
\begin{align}
\mathcal{T}^{\mu'}{}_{\nu'} = L^{\mu'}{}_\mu \mathcal{T}^\mu{}_\nu L^\nu{}_{\nu'} \qquad \text{and} \qquad D\mathcal{T}^{\mu'}{}_{\nu'} = L^{\mu'}{}_\mu \big( D\mathcal{T}^\mu{}_\nu \big) L^\nu{}_{\nu'} \, .
\end{align}

As shown above, the structure group of the coordinate frame and coframe is $GL(4,\mathbb{R})$, where this group acts on the frame/coframe canonically. In group algebraic language, there exists a standard right-action of $GL(4,\mathbb{R})$ on the fiber, which consists of frames/coframes in this context. Moreover, one may enrich the structure of the manifold by adding new objects (often coming with some constraints). By doing so, according to the new constraints on the manifold, the structure group also receives corresponding restrictions. Generally, the group reduces to a subgroup, $\mathcal{G'}\subset\mathcal{G}$. For instance, if we equip the manifold with a general Riemannian metric--which makes the manifold an inner product Riemannian one--it allows us to construct an orthogonal frame bundle over the Riemannian manifold. This process reduces the $GL(4,\mathbb{R})$ to $O(4)\subset GL(4,\mathbb{R})$. Furthermore, by imposing normalization and pseudo-Riemannian conditions as well, we may define the orthonormal frame bundle associated with the Lorentz group $SO(1,3)\subset GL(4,\mathbb{R})$, which allows us to construct the Lorentz metric locally with the help of the tetrad (vierbein) in the following way.

By means of the tetrad (vierbein) $h^a{}_\mu$, we lift $CT^*(M)$ to the orthonormal coframe bundle $OT^*(M) = U \times \mathbb{R}^{1,3}$ via $e^a = h^a{}_\mu dx^\mu$, where $e^a \in OT^*(M)$, $U \subset M$ is a region and $\mathbb{R}^{1,3}$ is a four-dimensional real vector space. Thus, $e^a$ is a section of this bundle. The metric is lifted from $CT^*(M)$ to $OT^*(M)$ by the inverse tetrad $h^\mu{}_a$ as $\eta_{ab} = h^\mu{}_a g_{\mu\nu} h^\nu{}_b$, where $\eta_{ab}$ are the Minkowski metric components and $h^a{}_\mu h^\mu{}_b = \delta^a_b$. Hence, in $OT^*(M)$ the metric reads $g=\eta_{ab} e^a \otimes e^b$. The connection 1-form is lifted according to
$\omega^a{}_b = h^a{}_\mu \omega^\mu{}_\nu h^\nu{}_b + h^a{}_\mu dh^\mu{}_b$. Two orthonormal coframes $e^{a}$ and $e^{a'}$ are related by $e^{a'} = L^{a'}{}_a e^a$. Requiring invariance of $g=\eta_{ab} e^a \otimes e^b = \eta_{a'b'} e^{a'} \otimes e^{b'}$ yields $\eta_{a'b'} = L^{a}{}_{a'} \eta_{ab} L^{b}{}_{b'}$, with $\eta_{a'b'} = \eta_{ab} = \mathrm{diag}(-1,+1,+1,+1)$. In matrix notation, this is $[\eta] = [L]^T [\eta] [L]$, implying that $L^{a'}{}_a$ forms the Lorentz group $SO(1,3)$. Thus, $SO(1,3)$ is the structure group of $OT^*(M)$. Moreover, the $(1,1)$-type tensor-valued exterior form living on $CT^*(M)$ and its covariant exterior derivative are lifted to $OT^*(M)$ as
$\mathcal{T}^a{}_b = h^a{}_\mu \mathcal{T}^\mu{}_\nu h^\nu{}_b$ and $D\mathcal{T}^a{}_b = h^a{}_\mu \big( D\mathcal{T}^\mu{}_\nu \big) h^\nu{}_b$, where
\begin{align}
D \mathcal{T}^a{}_b = d\mathcal{T}^a{}_b + \omega^a{}_c \wedge \mathcal{T}^c{}_b - \omega^c{}_b \wedge \mathcal{T}^a{}_c \, .
\end{align}
Then, under the transformation of the orthonormal coframes $e^{a'} = L^{a'}{}_a e^a$, both $\mathcal{T}^a{}_b$ and its covariant exterior derivative transform in the same manner:
\begin{align}
\mathcal{T}^{a'}{}_{b'} = L^{a'}{}_a \mathcal{T}^a{}_b L^b{}_{b'} \qquad \text{and} \qquad D\mathcal{T}^{a'}{}_{b'} = L^{a'}{}_a \big( D\mathcal{T}^a{}_b \big) L^b{}_{b'} \, .
\end{align}

Finally, by means of the Dirac matrices, we lift the entire structure from the $OT^*(M)$ bundle to the spinor bundle $\mathcal{S}(M)=U \times \mathbb{C}^4$, where $U \subset M$ is a region and $\mathbb{C}^4$ is a four-dimensional complex vector space. The complex vectors defined on this space are called spinors $\psi$. That is, the spinor $\psi$ is a four-component complex column matrix. Hence, each point on a fiber of $\mathcal{S}(M)$ represents a spinor. In other words, $\psi$ is a section of this bundle, and $\overline{\psi}:= \psi^\dagger \gamma_0$ is its projection, where the dagger symbol denotes Hermitian conjugation. An orthonormal coframe at a point of the $OT^*(M)$ bundle can be lifted to $\mathcal{S}(M)$ via the assignment $\gamma = \gamma_a e^a$. However, lifting the connection 1-form of the $OT^*(M)$ bundle to the spinor bundle is not as simple and canonical as lifting the coframe. For the moment, let us denote the connection 1-form of the spinor bundle by $\Omega$ and refer to it as the spinor connection 1-form. Accordingly, the covariant exterior derivative of the spinor is written as
\begin{align}
D\psi = d\psi + \Omega \psi \, .
\end{align}

Two spinors $\psi$ and $\psi'$ corresponding to two different points on a fiber of $\mathcal{S}(M)$ are related linearly by $\psi' = S \psi$, where $S$ is a four-component complex transformation matrix. This matrix $S$ is an element of the structure group of the spinor bundle, $S \in \mathcal{G}$. According to the covariantization rule, in order for $D\psi' = S (D\psi)$ to hold under the transformation rule $\psi' = S \psi$, the spinor connection 1-form must transform as
\begin{align}
\Omega' = S \Omega S^{-1} + S dS^{-1} \, . \label{eq:spin-connec-donusum}
\end{align}
At the same time, the transformation rule between orthonormal coframes in the spinor bundle can be written as $\gamma' = S \gamma S^{-1}$. In this case, each orthonormal coframe transformation is represented by two distinct elements, $S$ and $-S$, of $\mathcal{G}$, which act differently on spinor fields. For this reason, the group $\mathcal{G}$ is said to be a double cover of $SO(1,3)$.

For instance, in Einstein-Dirac theory, the lifting is achieved by the assignment $\Omega = \frac{1}{2} \sigma_{ab} \omega^{ab}$. Since in General Relativity the gravitational field is geometrized, $\Omega$ essentially represents the interaction of the spinor with the gravitational field. In this case, $\mathcal{G} = Spin(1,3)$. In Einstein-Dirac-Maxwell theory, one has $\Omega = \frac{1}{2} \sigma_{ab} \omega^{ab} + iqA$, where $q$ is the electric charge of the spinor and $A$ is the Maxwell potential 1-form. In this case, $\mathcal{G} = Spin(1,3) \otimes U(1)$. In Einstein-Cartan-Dirac-Maxwell theory, $\Omega = \frac{1}{2} \sigma_{ab} \omega^{ab} - \frac{1}{2} I T + iqA$, where $T$ is the torsion trace 1-form. In this case, $\mathcal{G} = Spin(1,3) \otimes W(1) \otimes U(1)$, where $W(1)$ denotes the Weyl group \cite{adak-ozdemir2023}. At this stage, it is worth emphasizing that $Spin(1,3)$, $Spin(1,3) \otimes U(1)$, and $Spin(1,3) \otimes W(1) \otimes U(1)$ are all double covers of $SO(1,3)$. Ultimately, the group $\mathcal{G}$ is larger than $SO(1,3)$ yet smaller than $GL(4,\mathbb{R})$. Moreover, depending on the interaction freedom of the spinor, the structure group of the spinor bundle enlarges. Similarly, for the novel covariant exterior derivative of the spinor proposed in (\ref{eq:covar-deriv-spinor-new1}), the structure group $\mathcal{G}$ is a double cover of the Lorentz group and remains smaller than $GL(4,\mathbb{R})$. The precise structure of this group has not yet been determined; it will be the subject of a separate future investigation.

One may further interpret the proposed novel covariant exterior derivative (\ref{eq:covar-deriv-spinor-new1}) within the framework of Koszul $\mathcal{G}$-connections. Recall that a Koszul $\mathcal{G}$-connection on a $\mathcal{G}$-structure is characterized by a covariant derivative whose associated connection 1-form takes values in the Lie algebra $\mathfrak{g}$ of the structure group $\mathcal{G}$. In local form, the transformation law (\ref{eq:spin-connec-donusum}) ensures precisely that $\Omega$ defines a principal $\mathcal{G}$-connection and hence a Koszul $\mathcal{G}$-connection on the associated vector bundle. From this perspective, the novel covariant exterior derivative (\ref{eq:covar-deriv-spinor-new1}) may be viewed as defining a new class of Koszul $\mathcal{G}$-connections on the spinor bundle $\mathcal{S}(M)$. In particular, if the corresponding connection 1-form $\Omega$ is $\mathfrak{g}$-valued for a structure group $\mathcal{G}$ that is a double cover of $SO(1,3)$ yet strictly smaller than $GL(4,\mathbb{R})$, then the resulting geometry corresponds to a reduced $\mathcal{G}$-structure endowed with a Koszul $\mathcal{G}$-connection that is not necessarily the lift of the Levi-Civita connection.

Consequently, our novel covariant derivative may generate a geometrically nontrivial extension of the standard spin connection: it defines a Koszul $\mathcal{G}$-connection compatible with the Clifford module structure while potentially allowing for generalized torsion, non-metricity, or additional internal gauge components encoded in $\mathfrak{g}$. In this sense, (\ref{eq:covar-deriv-spinor-new1}) does not merely modify the spinor covariant derivative at the level of field theory, but may correspond to a genuinely new class of reduced $\mathcal{G}$-geometries on $M$, whose precise algebraic and geometric characterization remains to be fully determined.

\section{Evolution of the Dirac equation} \label{sec:evolution-of-dirac-eqn}

Paul Dirac derived the relativistic wave equation for a free electron with a mass $m$ and a momentum $\vec{p}= -i\hbar (\partial/\partial x, \partial/\partial y, \partial/\partial z)$ in Cartesian coordinates in 1928 
  \begin{align}
      i\hbar \frac{\partial \psi}{\partial t} = \frac{\hbar c}{i} \left( \alpha_1  \frac{\partial \psi}{\partial x} + \alpha_2 \frac{\partial \psi}{\partial y} + \alpha_3 \frac{\partial \psi}{\partial z}\right) + \beta mc^2 \psi 
 \end{align}
where $\hbar = h/2\pi$ is the reduced Planck constant, $c$ is the speed of light in a vacuum, $\beta, \alpha_1,\alpha_2,\alpha_3$ are the Dirac matrices, and $\psi$ is the Dirac spinor \cite{dirac1928,bjorken1964}. This equation could be recast in the language of exterior forms as
  \begin{align}
      i*\gamma \wedge d\psi + i\frac{mc}{\hbar} \psi *1 =0 \label{eq:dirac-eqn-cart-coor}
  \end{align}
where $*$ (asterisk) is the Hodge map and $\gamma = \gamma_ae^a$ is a $cl(1,3)$-valued 1-form. The appearance of the imaginary unit $i$ depends on the choice of gamma matrices and the signature of the metric. The factor $i$ in both the kinetic and mass terms is required in our conventions for the Dirac Lagrangian to be Hermitian. The Hermiticity of the Dirac Lagrangian leads to a charge current which admits the usual probabilistic interpretation \cite{hehl-obukhov-1997}.

In addition to declaring the ``direct Dirac equation'', the Lagrangian formulation of the theory in the field theoretical approach is also a very powerful, effective, and widely accepted method. Thus, we will name the Dirac equation obtained by variation of the Lagrangian the ``variational Dirac equation'' in order to clarify and solidify our arguments in the following sections. Consequently, we write down the Dirac Lagrangian 4-form from which the direct Dirac equation given by Eq.~(\ref{eq:dirac-eqn-cart-coor}) is derived by varying with respect to $\overline{\psi}$ in natural units $\hbar = c =1$
   \begin{align}  \label{eq:dirac-lagrangian-cart1}
      L_D   = \frac{i}{2} \left[ \overline{\psi} * \gamma  \wedge  d\psi + d\overline{\psi} \wedge *\gamma  \psi \right] + im \overline{\psi} \psi *1
  \end{align}
where $\overline{\psi}:= \psi^\dagger \gamma_0$ is the Dirac adjoint spinor. As explained in the previous paragraph, the Dirac Lagrangian is Hermitian. The second term in this Lagrangian can be written as
\begin{eqnarray}
    d\overline{\psi} \wedge *\gamma \psi =   \overline{\psi}  *\gamma\wedge d\psi - \overline{\psi}  (d*\gamma) \psi   + d( \overline{\psi}  *\gamma \psi) .
\end{eqnarray}
Since $d*\gamma = 0$ in Cartesian coordinates and an exact form does not contribute to the variational field equations, $\overline{\psi}$-variation of the first and second terms yields the same result $\delta \overline{\psi}  *\gamma\wedge d\psi$. Thus, $\overline{\psi}$-variation of the Dirac Lagrangian gives rise to the variational Dirac equation, which is the same as the direct Dirac equation given by Eq.~(\ref{eq:dirac-eqn-cart-coor}). 

When the spin-1/2 fermion with mass $m$ has an electric charge $q$ and interacts with an external electromagnetic field, one can write down the direct Dirac equation by applying the minimal coupling recipe, which means replacing the exterior derivative $d$ with the covariant exterior derivative $\mathcal{D}$,
   \begin{align} 
      i*\gamma \wedge \mathcal{D}\psi + im \psi *1  = 0 \quad \text{where} \quad  \mathcal{D} \psi = d \psi + iqA \psi  \label{eq:dirac-eqn-U1} 
  \end{align}
and $A$ is the electromagnetic potential 1-form. By applying the minimal coupling prescription to the Dirac Lagrangian, we can write down the following:
 \begin{align}  \label{eq:qed-lagranjiyen}
      L_{DM}   = \frac{i}{2} \left[ \overline{\psi} * \gamma  \wedge  \mathcal{D}\psi + \mathcal{D}\overline{\psi} \wedge *\gamma  \psi \right] + im \overline{\psi} \psi *1 
  \end{align}
where $\mathcal{D} \overline{\psi} := \left(\mathcal{D} \psi \right)^\dagger \gamma_0 = d\overline{\psi} - iqA \overline{ \psi}$. The variational Dirac equation is obtained by varying this Lagrangian with respect to $\overline{\psi}$. It is observed that the direct Dirac equation and the variational Dirac equation are the same, as expected. 

Similarly, for a spin-1/2 fermion with mass $m$ interacting with a gravitational field represented by Einstein's theory of general relativity formulated in the Riemannian spacetime, one can write down the direct Dirac equation via the minimal coupling recipe, i.e., $d \to \widetilde{D}$,
    \begin{align}
      i*\gamma \wedge \widetilde{D}\psi + im \psi *1  = 0   \quad \text{where} \quad \widetilde{D}\psi = d\psi + \frac{1}{2}\widetilde{\omega}^{ab} \sigma_{ab}\psi . \label{eq:direct-dirac-eqn-gr}
  \end{align}
Again, the corresponding Lagrangian is cast readily,
    \begin{align}  \label{eq:dirac-lagran-gr}
      \widetilde{L}_{DE}   = \frac{i}{2} \left[ \overline{\psi} * \gamma  \wedge  \widetilde{D}\psi + \widetilde{D}\overline{\psi} \wedge *\gamma  \psi \right] + im \overline{\psi} \psi *1
  \end{align}
where $\widetilde{D} \overline{\psi} := \left(\widetilde{D} \psi \right)^\dagger \gamma_0 = d\overline{\psi} - \frac{1}{2}\widetilde{\omega}^{ab} \overline{\psi} \sigma_{ab}$. 
The second term in this Lagrangian can be written as 
  \begin{eqnarray}
     \widetilde{D}\overline{\psi} \wedge *\gamma  \psi   =  \overline{\psi} * \gamma  \wedge  \widetilde{D}\psi  - \overline{\psi}(\widetilde{D}*e^a - \widetilde{\omega}^{(ab)} \wedge*e_b)\gamma_a\psi + d(\overline{\psi} * \gamma  \wedge  \psi)\;.
  \end{eqnarray}
Since $\widetilde{D}*e^a=0$ and $\widetilde{\omega}^{(ab)}=0$ in Riemannian geometry,  the variational Dirac equation is easily obtained from the $\overline{\psi}$-variation of the Dirac Lagrangian. Thus, we find that the direct Dirac equation and the variational Dirac equation are the same, as expected. 

By following the same strategy in the literature, the direct Dirac equation in the Riemann-Cartan spacetime is written explicitly, 
 \begin{align}
      i*\gamma \wedge \widehat{D}\psi + im \psi *1  = 0 \quad \text{where} \quad  \widehat{D}\psi = d\psi + \frac{1}{2}\widehat{\omega}^{ab} \sigma_{ab}\psi \label{eq:direct-dirac-eqn-riemann-cartan} 
  \end{align}
and $\widehat{\omega}^{ab} = \widetilde{\omega}^{ab} + K^{ab}$. Then, the appropriate Dirac Lagrangian \cite{trautman1973,hehl-datta1971} turns out to be 
   \begin{align}  \label{eq:dirac-lagran-riemann-cartan}
      \widehat{L}_{DRC}   = \frac{i}{2} \left[ \overline{\psi} * \gamma  \wedge  \widehat{D}\psi + \widehat{D}\overline{\psi} \wedge *\gamma  \psi \right] + im \overline{\psi} \psi *1
  \end{align}
where $\widehat{D} \overline{\psi} := \left(\widehat{D} \psi \right)^\dagger \gamma_0 = d\overline{\psi} - \frac{1}{2}\widehat{\omega}^{ab} \overline{\psi} \sigma_{ab}$. However, the variational Dirac equation, which is derived by varying $ \widehat{L}_{DRC}$ with respect to $\overline{\psi}$,
   \begin{align}
     i  * \gamma  \wedge \left( \widehat{D}\psi - \frac{1}{2} T\psi \right) + im \psi *1 =0 \label{eq:dirac-eqn-varyas-riemann-cartan} 
 \end{align}
is, surprisingly, not the same as the direct Dirac equation given by Eq.~(\ref{eq:direct-dirac-eqn-riemann-cartan}) for the Riemann-Cartan spacetime. This extraordinary result, which does not manifest itself in Riemannian spacetime, has been extensively investigated with different motivations and names \cite{hehl-datta1971,obukhov-pereira2004,delhom2020,adak2012,adak-ozdemir2023}, and the references therein. However, at the same time, this distinction is not limited solely to Riemann-Cartan spacetime but is a general consequence of any metric-affine spacetime \cite{adak2023sce,vacaru2025,neeman1977,vacaru2004,adak2003,jimenez2020}, and the references therein. We call this discrepancy the {\it inconsistency problem} of the Dirac equation in non-Riemannian geometries. One remedy for it was discussed in three-dimensional Riemann-Cartan spacetime in Ref. \cite{adak-ozdemir2023} by extending the covariance group from the Lorentz group to the Lorentz-Weyl group. Although the Clifford algebra $cl(1,2)$ has the basis $\{I,\gamma_a,\sigma_{ab}\}$ in three dimensions, that discussion does not include the term $\gamma_a$ in the spinor connection. In addition, non-metricity is never even mentioned in the discussion. Thus, we will handle this inconsistency problem from as wide a perspective as possible in the following section by including all basis elements of the Clifford algebra $cl(1,3)$ in the spinor connection in four dimensions.

\section{Our resolution to the inconsistency problem} \label{sec:our-solution-to-inconsis}

In this paper, we discuss the inconsistency problem and propose a solution in a novel way in four-dimensional metric-affine spacetimes from the most general perspective. From the analysis presented so far, we conclude that the spinor bundle connection 1-form has a general structure 
 \[
   \Omega=\sum \left(\text{metric-affine connection 1-form}\right) \times \left(\text{certain basis elements of }cl(1,3)\right).
 \]
We argue $\Omega$ must include not only $\sigma_{ab}$, but also {\it all} the basis elements of $cl(1,3)$ and correspondingly all the possible ingredients of the geometry. Since the dynamics of a metric-affine geometry are characterized only by $e^a$ and $\omega^a{}_b$ in the orthonormal frames, by stating all the possible ingredients of the geometry, we mean both. Therefore, for the first time, we suggest
  \begin{subequations} \label{eq:covar-deriv-spinor-new1}
    \begin{align}
      \mathbb{D}\psi & = (d + \Omega) \psi \qquad \text{where}  \\
      \Omega  = \frac{1}{2} \sigma_{ab} \omega^{[ab]} + (a_1 I  + a_2 \gamma_5) Q &+ (a_3 I  + a_4 \gamma_5) P  
       + (b_1 I  + b_2 \gamma_5) T + (b_3 \gamma_a + b_4 \gamma_a\gamma_5) e^a \label{eq:generalized-connection}
  \end{align}
  \end{subequations}
where $\Omega$ is the spinor connection 1-form, while $a_i$ and $b_i$ ($i=1,2,3,4$) denote arbitrary complex coupling constants. Equivalently, the proposed formulation involves sixteen independent real coupling parameters. Its Dirac adjoint is obtained by using the identities given by Eq.~(\ref{eq:gamma0-GammaAdag-gamma0}) as
  \begin{align}
      \mathbb{D}\overline{\psi} &:=  \left(\mathbb{D} \psi \right)^\dagger \gamma_0  = d\overline{\psi} - \overline{\psi} (\gamma_0 \Omega^\dagger \gamma_0) \\
      &= d\overline{\psi} + \overline{\psi}\Big[- \frac{1}{2}\sigma_{ab}\omega^{[ab]} 
      + (a_1^\star I  + a_2^\star \gamma_5)Q + (a_3^\star I  + a_4^\star\gamma_5) P \nonumber\\  
      &\qquad \qquad \qquad  + (b_1^\star I  + b_2^\star \gamma_5)T - (b_3^\star \gamma_a + b_4^\star \gamma_5 \gamma_a) e^a \Big]
  \end{align}
where $^\star$ (star) denotes the complex conjugate. Accordingly, the direct Dirac equation for the most general non-Riemannian spacetime is written down
 \begin{align}
      i*\gamma \wedge \mathbb{D}\psi + im \psi *1  = 0  \;. \label{eq:direct-dirac-eqn-nonriemann} 
  \end{align}
Then, the corresponding Dirac Lagrangian becomes
   \begin{align}  \label{eq:dirac-lagran-nonriemann}
      {L}_{\mathbb{D}}   = \frac{i}{2} \left[ \overline{\psi} * \gamma  \wedge  \mathbb{D}\psi + \mathbb{D}\overline{\psi} \wedge *\gamma  \psi \right] + im \overline{\psi} \psi *1 .
  \end{align}
Now, we compute the variational Dirac equation by varying the Lagrangian $L_{\mathbb{D}}$ with respect to $\overline{\psi}$. For this purpose, we rewrite the second term in the Lagrangian as
  \begin{align}
 \mathbb{D}\overline{\psi} \wedge *\gamma  \psi&=   \overline{\psi} \wedge *\gamma \wedge \Big\{ d\psi + \frac{1}{2}  \sigma_{ab} 
      \omega^{[ab]}  \psi + (-a_1^\star I  + a_2^\star \gamma_5)Q \psi \nonumber \\
    &\quad
      + (-a_3^\star I  + a_4^\star\gamma_5) P   \psi 
      +(-b_1^\star I  + b_2^\star \gamma_5)T \psi   \nonumber + (b_3^\star I  - b_4^\star \gamma_5 ) \gamma  \psi \Big\}    \nonumber \\
    &\quad    - \overline{\psi}(D*e^a - \omega^{ab}\wedge*e_b)\gamma_a\psi + d(\overline{\psi} * \gamma  \wedge  \psi) .
\end{align}
In this step, the commutator relations given by Eq.~(\ref{eq:commut-rel-cl-basis}) have been used repeatedly. Besides, we insert the following result derived from the identity in Eq.~(\ref{eq:Dhodgeea}) into the expression above 
   \begin{eqnarray}
         D*e^a - \omega^{(ab)}\wedge*e_b =*e_a  \wedge (Q+T-P)  .
   \end{eqnarray}
Then we arrive at
    \begin{align}
 \mathbb{D}\overline{\psi} \wedge *\gamma  \psi &=   \overline{\psi} \wedge *\gamma \wedge \Big\{ d + \frac{1}{2}  \sigma_{ab} 
      \omega^{[ab]}   + [-(a_1^\star +1)I  + a_2^\star \gamma_5]Q 
      + [-(a_3^\star-1) I  \nonumber \\
    &\quad+ a_4^\star\gamma_5] P    
      +[-(b_1^\star +1)I + b_2^\star \gamma_5]T + (b_3^\star I  - b_4^\star \gamma_5 ) \gamma   \Big\} \psi + d(\overline{\psi} * \gamma  \wedge  \psi) .
 \end{align}
Meanwhile, we explicitly rewrite the first term in the Lagrangian given by Eq.~(\ref{eq:dirac-lagran-nonriemann}) as
 \begin{align}
     \overline{\psi} \wedge *\gamma \wedge \mathbb{D}\psi & = \overline{\psi} \wedge *\gamma \wedge \Big\{ d + \frac{1}{2} \sigma_{ab} \omega^{[ab]} + (a_1 I  + a_2 \gamma_5) Q + (a_3 I  + a_4 \gamma_5) P  \nonumber \\
      & \qquad  + (b_1 I  + b_2 \gamma_5) T + (b_3 I - b_4 \gamma_5) \gamma    \Big\} \psi .
  \end{align}
Thus, $\overline{\psi}$-variation of the Dirac Lagrangian yields the variational Dirac equation 
\begin{align}
  i*\gamma \wedge \Big\{ d + \frac{1}{2}  \sigma_{ab} \omega^{[ab]}  +\frac{1}{2}[ (a_1- a_1^\star-1) I  + (a_2+a_2^\star) \gamma_5]Q & \nonumber\\
     + \frac{1}{2}[(a_3-a_3^\star+1) I  + (a_4+a_4^\star)\gamma_5] P  
     +\frac{1}{2}[(b_1-b_1^\star-1) I  + (b_2+b_2^\star) \gamma_5]T  &  \nonumber \\
    + \frac{1}{2}[(b_3+b_3^\star)I -(b_4+b_4^\star) \gamma_5 ] \gamma   \Big\}  \psi   + im\psi*1 &= 0    \;. \label{eq:var-Dirac-eq}
\end{align} 
In order for this variational Dirac equation to be the same as the direct Dirac equation given by Eq.~(\ref{eq:direct-dirac-eqn-nonriemann}) the coupling constants must satisfy
  \begin{subequations}\label{eq:coeffs-constraints}
 \begin{align}
   a_1-a_1^\star-1 &= 2a_1, & a_2^\star &= a_2, &
    a_3-a_3^\star +1 &= 2a_3, & a_4^\star &= a_4, \\
   b_1-b_1^\star -1&= 2b_1,  & b_2^\star &= b_2, &
    b_3^\star &= b_3, & b_4^\star &= b_4 .
 \end{align}
  \end{subequations}
Under these choices the coupling coefficients could be written in the following form 
  \begin{subequations} \label{eq:Dirac Lagrangian varyasyon10_1}
    \begin{align}
    a_1&= -\frac{1}{2} +iA_1, & a_2 &= A_2, &
    a_3&= +\frac{1}{2} +iA_3, & a_4 &= A_4, \\
    b_1&= -\frac{1}{2} +iB_1, & b_2 &= B_2, &
    b_3&= B_3, & b_4 &=B_4, 
   \end{align}
  \end{subequations}
where $A_1,A_2,A_3,A_4,B_1,B_2,B_3,B_4\in\mathbb{R}$. 

The requirement that the variational and direct formulations of the Dirac equation coincide imposes eight independent constraints on the most general spinor connection, thereby reducing the number of independent real coupling constants from sixteen to eight. This reduction is not introduced ad hoc, but instead follows naturally from an internal consistency condition of the theory. Consequently, the resulting framework is substantially more constrained than its most general algebraic form while preserving its generality. From a broader perspective, this simplification is also in accordance with the spirit of Occam's razor, according to which theories with fewer independent assumptions are generally preferred when they possess comparable explanatory power. The equivalence of the variational and direct formulations of the Dirac equation may therefore be viewed not only as a consistency requirement but also as a physically motivated principle for selecting admissible spinor connections.

These consistency conditions also have a physical interpretation in terms of the Hermiticity of the generalized Dirac Hamiltonian. In the Foldy--Wouthuysen analysis of the symmetric teleparallel sector \cite{ugur_adak_bagci_pala_2026}, curvature and torsion vanish, while the non-metricity couplings remain. The historical review on the formative years and later developments of symmetric teleparallel gravity can be read in \cite{Adak2026-history-stpg} and the references therein. For a static isotropic metric $ds^2=-f^2(r)dt^2+g^2(r)\delta_{ij}dx^i dx^j$, the natural Dirac inner product on a spatial hypersurface $\Sigma_t$ and the corresponding formal adjoint of $p_i=-i\partial_i$ are, in the natural units used here,
 \begin{align}
  \langle\phi,\psi\rangle_{\Sigma_t}
  =\int_{\Sigma_t}d^3x\,g^3(r)\phi^\dagger\psi,
  \qquad
  p_i^\ddagger=p_i-i\partial_i\ln g^3,
 \end{align}
where $\ddagger$ denotes the adjoint with respect to this weighted inner product and boundary terms are assumed to vanish. The position-dependent kinetic term and the spatial measure must therefore be included when testing Hermiticity; testing only the algebraic matrix factors is insufficient.

For general static isotropic metric functions $f$ and $g$, formal Hermiticity of the complete relativistic Hamiltonian recovers
 \begin{align}
  \operatorname{Re}a_1=-\frac{1}{2},\qquad
  \operatorname{Re}a_3=\frac{1}{2},\qquad
  a_2,a_4,b_3,b_4\in\mathbb{R},
 \end{align}
in agreement with the corresponding conditions in Eq.~(\ref{eq:Dirac Lagrangian varyasyon10_1}). The fixed real parts of $a_1$ and $a_3$ compensate the adjoint contribution of the kinetic operator and the curved spatial measure, while their imaginary parts remain unrestricted by Hermiticity. Thus the Hamiltonian analysis provides a physical check, within the symmetric teleparallel sector, of the admissible couplings selected by direct--variational agreement. Since $T^a=0$ in Ref.~\cite{ugur_adak_bagci_pala_2026}, it does not independently constrain the torsion coefficients $b_1$ and $b_2$. The agreement in the surviving sector strengthens the interpretation of the consistency conditions as compatible with probability-conserving fermionic evolution.

\section{Effective fermion mass and chiral structure} \label{sec:effective mass}

The coframe-dependent terms in the generalized spinor connection
contribute to the fermionic mass sector even in Minkowski spacetime.
To isolate these contributions and clarify their chiral interpretation,
we rewrite the Dirac Lagrangian in Eq.~(\ref{eq:dirac-lagran-nonriemann}) in the following Hermitian form:
   \begin{align}
      L_{\mathbb{D}} &= \mathcal{H}er \left[ i\overline{\psi} * \gamma \wedge \mathbb{D}\psi + im \overline{\psi} \psi *1 \right] . \label{eq:summary-lagran-a}
    \end{align}
The Hermitian part of an expression $\mathcal{O}$ is defined by $\mathcal{H}er[\mathcal{O}] = (\mathcal{O} + \mathcal{O}^\dagger)/2$. Then we separate the coframe terms and the connection terms in the definition $\mathbb{D}\psi$ given by Eq.~(\ref{eq:covar-deriv-spinor-new1}) as follows
   \begin{align}
     \mathbb{D} \psi  = \widehat{\mathbb{D}} \psi + (b_3 \gamma_a + b_4 \gamma_a \gamma_5) e^a \psi
  \end{align}
where
   \begin{align}
     \widehat{\mathbb{D}} \psi := \left\{ d + \frac{1}{2} \sigma_{ab} \omega^{[ab]} + (a_1 I + a_2 \gamma_5) Q + (a_3 I + a_4 \gamma_5) P + (b_1 I + b_2 \gamma_5) T \right\} \psi .
  \end{align}
Thus, the first term in Eq.~(\ref{eq:summary-lagran-a}) becomes
 \begin{align}
  i\overline{\psi} * \gamma \wedge \mathbb{D}\psi = i\overline{\psi} * \gamma \wedge \widehat{\mathbb{D}}\psi - 4i\overline{\psi} (b_3I + b_4 \gamma_5) \psi *1
\end{align}
where we used the identity $ *e^b \wedge e^a = -\eta^{ab}*1$. We now substitute this result back into Eq.(\ref{eq:summary-lagran-a}):
  \begin{align}
    L_{\mathbb{D}} = \mathcal{H}er \left\{ i\overline{\psi} * \gamma \wedge \widehat{\mathbb{D}}\psi + i\overline{\psi} \left[ (m -4 b_3)I -4 b_4 \gamma_5\right] \psi *1 \right\}
    \label{eq:Dirac-lagran-decomposed-Her}
  \end{align}

Equation (\ref{eq:Dirac-lagran-decomposed-Her}) shows that the coframe-dependent terms in the generalized spinor connection in Eq.~(\ref{eq:generalized-connection}) introduce scalar and pseudoscalar contributions to the fermionic mass sector. Importantly, these contributions survive in Minkowski spacetime. Indeed, for an inertial orthonormal coframe $e^a=dx^a$ and $\omega^a{}_b=0$, curvature, torsion, and non-metricity vanish, while the terms $(b_3\gamma_a+b_4\gamma_a\gamma_5)e^a$ remain in the generalized spinor connection. Thus, the mass structure obtained in Eq.~(\ref{eq:Dirac-lagran-decomposed-Her}) does not require a non-Riemannian background. With the real coefficients selected by the consistency conditions, we define
 \begin{align}
  m_s:=m-4b_3,\qquad
  m_p:=4b_4,\qquad
  \mathsf{M}:=m_s I - m_p \gamma_5.
 \end{align}
Here $b_3$ modifies the scalar mass coefficient, whereas $b_4$ introduces a pseudoscalar mass coefficient. Their inclusion in the generalized connection provides a geometrical representation of this extended mass structure, including its flat-spacetime limit.

The persistence of the $b_3$ and $b_4$ contributions in Minkowski
spacetime can also be seen directly at the level of the Dirac equation.
Under the consistency conditions in
Eq.~(\ref{eq:coeffs-constraints}), the variational Dirac equation
given by Eq.~(\ref{eq:var-Dirac-eq}) can be written in terms of
the extended mass operator introduced above as
\begin{align}
    i*\gamma\wedge\widehat{\mathbb{D}}\psi
    + i\mathsf{M}\psi*1 = 0.
\end{align}
In Minkowski spacetime, with an inertial orthonormal coframe
$e^a=dx^a$ and a vanishing affine connection $\omega^a{}_b=0$,
the reduced spinor covariant derivative becomes the ordinary
exterior derivative, $\widehat{\mathbb{D}}\psi=d\psi$, yielding
\begin{align}
    i*\gamma\wedge d\psi + i\mathsf{M}\psi*1 = 0.
\end{align}
Thus, the $b_3$ and $b_4$ contributions survive in this flat
background through $\mathsf{M}=(m-4b_3)I-4b_4\gamma_5$.
The resulting equation retains the standard Minkowski Dirac
kinetic operator, while the extended mass operator incorporates
a scalar mass shift controlled by $b_3$ and a pseudoscalar mass
contribution controlled by $b_4$.

The chiral interpretation of these terms requires attention to our
Clifford algebra conventions. Since $\gamma_5^2=-I$, we introduce
the normalized chirality operator and the corresponding projectors,
  \begin{align}
  \Gamma_{\mathrm{ch}}:=-i\gamma_5,\qquad
  P_{R,L}:=\frac12\left(I\pm\Gamma_{\mathrm{ch}}\right).
  \end{align}
The mass operator can then be written as
  \begin{align}
  \mathsf{M}
  =(m_s-im_p)P_R+(m_s+im_p)P_L.
  \end{align}
Consequently, the coefficients associated with the two chiral
components are complex conjugates. They describe a chiral phase
in a scalar--pseudoscalar Dirac mass term, rather than two independent
real rest masses. The Dirac mass term couples the left- and
right-chiral components of the same fermionic field.

For a single free Dirac field in Minkowski spacetime, this
interpretation can be checked directly using the Hamiltonian.
With $\beta=i\gamma_0$, $\alpha^i=\gamma_0\gamma_i$, and
$\mathbf{p}=-i\boldsymbol{\nabla}$, we obtain
   \begin{align}
  H=\boldsymbol{\alpha}\cdot\mathbf{p}
    +\beta(m_s I-m_p\gamma_5),
  \qquad
  H^2=\left(\mathbf{p}^2+m_s^2+m_p^2\right)I.
   \end{align}
The resulting dispersion relation therefore contains a single
effective mass,
  \begin{align}
  m_{\mathrm{eff}}
  =\sqrt{m_s^2+m_p^2}
  =\sqrt{(m-4b_3)^2+16b_4^2}.
  \end{align}
Even if the original mass parameter vanishes, the generalized
connection admits a nonzero effective mass,
  \begin{align}
  m=0
  \quad\Longrightarrow\quad
  m_{\mathrm{eff}}=4\sqrt{b_3^2+b_4^2}.
  \end{align}
This establishes an effective mass contribution that persists in
flat spacetime. A dynamical explanation of its origin would
additionally require determining how the coupling coefficients
arise and acquire their values.

For constant coefficients and $m_{\mathrm{eff}}\neq0$, the mass
operator may equivalently be expressed as
  \begin{align}
  \mathsf{M}=m_{\mathrm{eff}}e^{-\theta\gamma_5},
  \qquad
  \cos\theta=\frac{m_s}{m_{\mathrm{eff}}},
  \qquad
  \sin\theta=\frac{m_p}{m_{\mathrm{eff}}}.
  \end{align}
The constant chiral field redefinition (or global chiral transformation) 
  \begin{align}
  \psi=e^{\theta\gamma_5/2}\chi,
  \qquad
  \overline{\psi}
  =\overline{\chi}e^{\theta\gamma_5/2}
  \end{align}
preserves the kinetic term and converts the mass contribution into
$im_{\mathrm{eff}}\overline{\chi}\chi\,*1$. Thus, the free
Minkowski-space dynamics depends on $m_{\mathrm{eff}}$, while the
scalar--pseudoscalar phase can be removed by a field redefinition.
A closely related construction appears in \cite{adak2004}, where
dimensional reduction produces an extended fermionic mass structure
and a chiral rotation identifies an effective mass. In that setting,
the rotation is generally position dependent and also induces
derivative interactions. This provides a direct connection with
our result through the interplay between effective mass, chiral
rotations, and additional fermionic couplings.

The relation between fermion mass and handedness has also been
investigated in the context of weak interactions. Franson
\cite{franson2025} discusses charged-current amplitudes involving
massive fermions and argues that contributions associated with
right-handed states can become larger than those associated with
left-handed states for sufficiently large masses, as illustrated
by the top quark. The handedness terminology in that work refers
to helicity states, which approximate chirality eigenstates in
the relativistic limit. Similarly, Shi \cite{zhi-qiang-shi-2013}
investigates the polarization dependence of the weak-interaction
mass and reports a larger contribution for left-handed polarized
fermions than for right-handed ones moving with the same velocity.
These studies are relevant to the present discussion because they
examine how mass contributions and interaction responses depend
on fermionic handedness. Their interaction-dependent results
should be distinguished from the single rest mass obtained above
for a free Dirac field. A quantitative connection with our
generalized couplings requires incorporating the relevant
interactions and transforming all terms consistently under
the chiral field redefinition.

On the experimental side, the measurements reported in
\cite{abe-et-al1993,abe-et-al1994,abe-et-al2000} establish the
left--right cross-section asymmetry in $Z$ boson production
by $e^+e^-$ collisions with progressively improved precision.
These measurements demonstrate the physical significance of
polarization-dependent fermionic interactions and provide
relevant observables for investigating extensions of the
fermionic sector. Their established electroweak interpretation
involves the vector and axial-vector couplings of the $Z$ boson
to the electron current. To connect the mass structure in
Eq.~(\ref{eq:Dirac-lagran-decomposed-Her}) with such observables, one must calculate the
corresponding interaction amplitudes, account for the allowed
field redefinitions, and identify any contribution that remains
physically distinguishable. Since the $b_3$ and $b_4$ terms
already survive in Minkowski spacetime, their possible effects
should be investigated as consequences of the generalized
fermionic couplings, without presupposing a gravitational
origin for the measured asymmetry.


Returning to the Lagrangian formulation, we evaluate the Hermitian part in Eq.~(\ref{eq:Dirac-lagran-decomposed-Her}) under the consistency conditions given in Eq.~(\ref{eq:coeffs-constraints}), obtaining
 \begin{align} \label{eq:dirac-lag-asymmetric}
    L_{\mathbb{D}} = i\overline{\psi} * \gamma \wedge \widehat{\mathbb{D}}\psi + i\overline{\psi}\left[(m -4 b_3)I -4b_4\gamma_5 \right]\psi *1 + \frac{i}{2} d(\overline{\psi}*\gamma\wedge\psi). 
 \end{align}
The last term is an exact four-form and does not contribute to the bulk variational field equations when the spinor variations vanish on the boundary. Under the consistency conditions in Eq.~(\ref{eq:coeffs-constraints}), the unsymmetrized form in Eq.~(\ref{eq:dirac-lag-asymmetric}) is equivalent to the manifestly Hermitian, symmetrized form in Eq.~(\ref{eq:Dirac-lagran-decomposed-Her}) up to an exact four-form, and allows the Dirac operator acting on \(\psi\), including its scalar and pseudoscalar mass contributions, to be read off directly. Although Hermiticity is no longer manifest in the bulk Lagrangian density, both forms yield the same bulk field equations when the boundary variation vanishes.


\section{Curvature of the generalized spinor connection}
\label{sec:spinor-curvature}

The analysis in Section~6 shows that the coframe-dependent
couplings survive in Minkowski spacetime. We now compute the
curvature of the full generalized spinor connection given in
Eq.~(\ref{eq:generalized-connection}) and distinguish it from
the curvature of the underlying affine connection. Throughout
this calculation, the coupling coefficients are constant complex
numbers; the consistency conditions on these coefficients are
not required for the derivation. The gamma matrices are constant
in the chosen local orthonormal trivialization.

Extending $\mathbb{D}=d+\Omega$ to spinor-valued differential forms,
the graded Leibniz rule gives
\begin{align}
  \mathbb{D}(\mathbb{D}\psi) = d(d\psi+\Omega\psi)
     +\Omega\wedge(d\psi+\Omega\psi) = (d\Omega+\Omega\wedge\Omega) \psi.
  \label{eq:spinor-curvature-definition}
\end{align}
Here $\mathcal{R} := d\Omega+\Omega\wedge\Omega$ is the endomorphism-valued curvature 2-form of the generalized spinor connection. To organize the calculation, we introduce
\begin{align}
  \Lambda_{ab}&:=\omega_{[ab]},
  & S&:=\frac{1}{2}\sigma_{ab} \Lambda^{ab}, & \Gamma_a&:=b_3\gamma_a+b_4\gamma_a\gamma_5, \nonumber \\
  \mathcal{A}&:=a_1Q+a_3P+b_1T,
  & \mathcal{B}&:=a_2Q+a_4P+b_2T .
\end{align}
Since $\sigma_{ab}$ is antisymmetric, the generalized connection (\ref{eq:generalized-connection}) can then be written as
 \begin{align}
  \Omega=S+I\mathcal{A}
             +\gamma_5\mathcal{B}
             +\Gamma_a e^a .
  \end{align}

The identity matrix commutes with all Clifford basis elements,
and $\gamma_5$ commutes with $\sigma_{ab}$. Consequently, the
cross terms involving $I\mathcal{A}$ cancel by the antisymmetry
of the exterior product, as do the cross terms between $S$ and
$\gamma_5\mathcal{B}$. Moreover, $ \mathcal{A}\wedge\mathcal{A}
  =\mathcal{B}\wedge\mathcal{B}=0$. Using the Clifford commutation relations, we obtain
\begin{align}
  [S,\Gamma_a]
  &=\Lambda^b{}_a\Gamma_b, & 
  [\Gamma_a,\Gamma_b]
  &=4(b_3^2+b_4^2)\sigma_{ab}, &
  [\gamma_5,\Gamma_a]
  &=2b_4\gamma_a-2b_3\gamma_a\gamma_5 .
\end{align}
The positive sign between $b_3^2$ and $b_4^2$ follows from our
convention $\gamma_5=\gamma_0\gamma_1\gamma_2\gamma_3$, for which
$\gamma_5^2=-I$. Combining these results yields
\begin{align}
  \mathcal{R}
  ={}&dS+S\wedge S
      +I\,d\mathcal{A}
      +\gamma_5\,d\mathcal{B} +\Gamma_a
       \left(de^a + \Lambda^a{}_b\wedge e^b\right) \nonumber \\
     &+2(b_3^2+b_4^2)\sigma_{ab}\,
       e^a\wedge e^b +2(b_4\gamma_a-b_3\gamma_a\gamma_5)\,
       \mathcal{B}\wedge e^a .
\end{align}
The decomposition $\omega^a{}_b = \Lambda^a{}_b+Q^a{}_b$ and the
Cartan structure equations imply
  \begin{subequations}
      \begin{align}
  R^{[ab]}
  &=d\Lambda^{ab} + \Lambda^a{}_c \wedge \Lambda^{cb}
    +Q^a{}_c\wedge Q^{cb},\\
  dS+S\wedge S
  &=\frac{1}{2}\sigma_{ab}
    \left(R^{[ab]}-Q^a{}_c\wedge Q^{cb}\right),\\
  de^a + \Lambda^a{}_b\wedge e^b
  &=T^a-Q^a{}_b\wedge e^b .
   \end{align}
  \end{subequations}
Furthermore, since $Q=\omega^a{}_a$ and
$\omega^a{}_b\wedge\omega^b{}_a=0$, we have $dQ=R^a{}_a$. Therefore, the curvature of the full generalized spinor
connection, decomposed over the Clifford basis, is
 \begin{align}
  \mathcal{R}={}&
  \frac{1}{2}\sigma_{ab}
  \left[
    R^{[ab]}
    -Q^a{}_c\wedge Q^{cb}
    +4(b_3^2+b_4^2)e^a\wedge e^b
  \right] \nonumber \\
  &+I\left(
    a_1R^a{}_a+a_3\,dP+b_1\,dT
  \right) + \gamma_5\left(
    a_2R^a{}_a+a_4\,dP+b_2\,dT
  \right) \nonumber \\
  &+\gamma_a\left[
    b_3\left(T^a-Q^a{}_b\wedge e^b\right)
    +2b_4(a_2Q+a_4P+b_2T)\wedge e^a
  \right] \nonumber \\
  &+\gamma_a\gamma_5\left[
    b_4\left(T^a-Q^a{}_b\wedge e^b\right)
    -2b_3(a_2Q+a_4P+b_2T)\wedge e^a
  \right]. \label{eq:spinor-curvature-full}
\end{align}
The combination $b_3^2+b_4^2$ in Eq.~(\ref{eq:spinor-curvature-full}) denotes the sum of the algebraic squares of the complex coefficients, rather than $|b_3|^2+|b_4|^2$. No complex conjugation occurs in this calculation, since the curvature is constructed directly from $\Omega$ without taking its Hermitian conjugate.

Two limiting cases clarify the geometrical content of this
result. For a torsion-free, metric-compatible connection with
$b_3=b_4=0$, Eq.~(\ref{eq:spinor-curvature-full}) reduces to
the standard spinor curvature,
 \begin{align}
  \mathcal{R}=\frac12\sigma_{ab}R^{ab}.
 \end{align}
In contrast, in Minkowski spacetime with an inertial orthonormal
coframe, the conditions $R^a{}_b=T^a=Q_{ab}=0$ give
 \begin{align}
  \mathcal{R}
  =2(b_3^2+b_4^2)\sigma_{ab}e^a\wedge e^b.
 \end{align}
Thus, flatness of the affine connection does not generally imply
flatness of the generalized spinor connection. The surviving
curvature originates from the noncommuting coframe-dependent
terms in $\Omega$. For the real coefficients selected by the
consistency conditions, it vanishes in this Minkowski limit
if and only if $b_3=b_4=0$.

This result also connects the curvature calculation with the
effective mass obtained in Section~\ref{sec:effective mass}. When the original mass
parameter vanishes, $m=0$, the relation
$m_{\mathrm{eff}}^2=16(b_3^2+b_4^2)$ yields
 \begin{align}
  \mathcal{R}
  =\frac{m_{\mathrm{eff}}^2}{8}
    \sigma_{ab}e^a\wedge e^b
  \qquad (m=0).
  \end{align}
The same combination of coframe couplings therefore determines
both the effective mass squared and the surviving spinor
curvature in this limit. This algebraic correspondence identifies
a common origin within the generalized connection, while the
underlying affine spacetime remains flat.

\section{Conclusions} \label{sec:discussion}

We have developed an exterior-algebra formulation of the
generalized Dirac equation in four-dimensional metric-affine
spacetimes. Comparing direct covariantization of the free
Dirac equation with variation of the covariantized,
symmetrized Lagrangian reveals geometric terms that prevent
the two procedures from agreeing for an unrestricted
spinor connection. The generalized connection in
Eq.~(\ref{eq:generalized-connection}) incorporates the
complete Clifford basis together with the antisymmetric
affine connection, non-metricity traces, the torsion trace,
and coframe-dependent couplings. Requiring the direct and
variational equations to coincide imposes eight independent
real constraints and reduces the initial sixteen real
coupling parameters to eight. This requirement provides
a criterion for selecting variationally consistent
connections within the proposed framework.

The Hamiltonian analysis gives a physical check of this
selection criterion in the static isotropic symmetric
teleparallel sector \cite{ugur_adak_bagci_pala_2026}.
Formal Hermiticity of the complete relativistic Hamiltonian,
tested with the curved-space Dirac inner product and the
appropriate adjoint of its position-dependent kinetic
operator, reproduces the corresponding restrictions on
the non-metricity and coframe couplings. The agreement
supports the compatibility of the consistency conditions
with probability-conserving fermionic evolution. Since
torsion vanishes in this sector, this check does not
independently determine the restrictions on the torsion
couplings.

The coframe-dependent terms introduce the extended mass
operator $\mathsf{M}=(m-4b_3)I-4b_4\gamma_5$, containing
scalar and pseudoscalar contributions that persist in
Minkowski spacetime. For a single free Dirac field with
constant coefficients, the dispersion relation yields
the single effective mass
$m_{\mathrm{eff}}=\sqrt{(m-4b_3)^2+16b_4^2}$.
The coefficients associated with the two chiral
components are complex conjugates and encode a chiral
phase, which can be removed by a constant chiral field
redefinition in this free limit. Their physical
interpretation therefore involves one rest mass and
a scalar--pseudoscalar phase. In particular, the
coframe couplings allow $m_{\mathrm{eff}}\neq0$ even
when $m=0$. A dynamical account of this mass requires
determining the origin and values of these couplings.

Under the consistency conditions, the symmetrized
Lagrangian can also be written in an unsymmetrized
form with an explicit exact four-form, as shown in
Eq.~(\ref{eq:dirac-lag-asymmetric}). This representation
makes the Dirac operator and its extended mass terms
directly accessible while preserving the bulk field
equations when the boundary variation vanishes.

The curvature calculation establishes a further
distinction between the generalized spinor connection
and the underlying affine connection. The
endomorphism-valued curvature
$\mathcal{R}=d\Omega+\Omega\wedge\Omega$ was obtained
for constant complex coefficients without imposing
the consistency conditions. Its Clifford decomposition,
given in Eq.~(\ref{eq:spinor-curvature-full}), exhibits
contributions from affine curvature, torsion,
non-metricity, and the additional spinor couplings.
In the Minkowski limit, the noncommuting coframe
terms yield
$\mathcal{R}=2(b_3^2+b_4^2)\sigma_{ab}e^a\wedge e^b$.
For the real $b_3$ and $b_4$ selected by consistency,
this curvature vanishes only when both coefficients
vanish. When $m=0$, it can equivalently be written as
$\mathcal{R}=(m_{\mathrm{eff}}^2/8)\sigma_{ab}e^a\wedge e^b$.
Thus, the effective mass squared and the surviving
spinor curvature share an algebraic origin in the
coframe couplings, even though the affine spacetime
remains flat.

The application to solar neutrino oscillations in
Ref.~\cite{cetinkaya_adak_2026} illustrates how the
generalized framework can be used to obtain
phenomenological constraints on its couplings.
Further investigations should extend the Hermiticity
analysis to torsionful backgrounds, determine the
covariance group of the generalized spinor connection,
and incorporate interactions and coupled gravitational
dynamics. In particular, assessing observable effects
of the extended mass structure requires transforming
all interaction terms consistently under the chiral
field redefinition.




 \bigskip 
\noindent 
{\bf Acknowledgements.} This work has been supported by TÜBİTAK (The Scientific and Technological Research Council of Türkiye) via the project 124F325. MA thanks Tomi Koivisto for fruitful discussions at the workshop of teleparallel universes in Salamanca, Spain, 26-28 November 2018. 

  \bigskip 
 \noindent
{\bf Data Availability Statement.} Data sharing is not applicable to this article as no datasets were generated or analyzed during the current theoretical research.

 \bigskip
 \noindent
{\bf Conflict of Interest.} The authors declare no conﬂict of interest.

\bigskip
 \noindent
\textbf{Declaration of generative AI.}
During the preparation of this manuscript, the authors used
ChatGPT (OpenAI) to assist with language editing, text organization,
and selected mathematical derivations. The authors reviewed and
revised the AI-assisted material and independently checked the
mathematical results. The authors take full responsibility for
the accuracy, originality, and scientific content of the manuscript.

\end{document}